\documentclass{aa} 

\newcommand{\filter}[1]{\mbox{\it #1\/}}              

\usepackage{graphicx}
\usepackage{txfonts}
\begin{document}

   \title{The very low-mass stellar content of the young supermassive Galactic star cluster Westerlund 1}

   \author{M. Andersen
          \inst{1}
          \and
          M. Gennaro \inst{2,3}
          \and 
          W. Brandner \inst{3}
          \and 
          A. Stolte \inst{4}
          \and 
          G. de Marchi \inst{5} 
          \and 
          M.~R. Meyer \inst{6}
          \and 
          H. Zinnecker \inst{7}
          }

   \institute{Gemini Observatory, Casilla 603, La Serena, Chile, \email{manderse@gemini.edu} 
         \and
             Space Telescope Science Institute
3700 San Martin Dr  Baltimore, MD 21218, USA 
          \and 
          Max-Planck-Institut f\"ur Astronomie, K\"onigstuhl 17, 69117 Heidelberg, Germany 
          \and
          Argelander Institut f\"ur Astronomie, Auf dem H\"ugel 71, 53121 Bonn, Germany
          \and 
          Research \& Scientific Support Department, ESA ESTEC, Keplerlaan 1, 2200 AG Noordwijk, The Netherlands
          \and 
          Institute for Astronomy ETH, Physics Department, HIT J 22.4, CH-8093 Zurich, Switzerland
          \and
          SOFIA-USRA, NASA Ames Research Center, MS N211-3, Moffett Field, CA 94035, USA
             }


  \abstract{
 We present deep near-infrared  HST/WFC3 observations of the young supermassive Galactic star cluster Westerlund 1 and an adjacent control field. 
 The depth of the data is sufficient to derive the mass function for the cluster as a function of radius down to 0.15 M$_\odot$ in the outer parts of the cluster.
  We identify 
for the first time a flattening  in the mass function 
(in logarithmic units) at a mass range that is consistent with that of the 
field and nearby embedded clusters. 
 Through log-normal functional fits  to the mass functions we find the nominal  peak mass to be comparable to that of the field and nearby embedded star clusters. 
The width of a log-normal fit appears slightly narrow compared to the width of the field IMF, closer to the values found for globular clusters. 
The subsolar content within the cluster does not appear to be mass segregated in contrast to the findings for the  supersolar content.
The total mass of Westerlund 1 is estimated to be  44-57 $\times 10^3$ M$_\odot$ where the main uncertainty is the choice of the isochrone age and the higher mass slope. 
Comparing the photometric mass with the dynamically determined  mass, Westerlund 1  is  sufficiently massive to remain bound and could potentially evolve into a low-mass globular cluster.}

   \keywords{Stars: low-mass -- 
     mass function  -- 
     (Galaxy:) open clusters and associations: individual: Westerlund 1 -- 
      Stars: formation
               }

   \maketitle
%

\section{Introduction}
The initial mass function (IMF) describes the distribution of stellar and brown dwarf masses for a star formation event and is thus a key quantity in many fields of astronomy, including interpreting the integrated light from unresolved stellar populations, Galactic models, and the evolution of star clusters. 
One important question is whether the IMF is dependent of environment or if it is universal { with a shape similar to that of the Galactic field \citep{chabrier,kroupa_new} with a general flattening below 1 M$_\odot$.} 
Extensive work has been carried out in this regard for the nearby lower mass clusters where there has been found little to no evidence of variations \citep[e.g.][]{andersen08,bastian,kroupa_new,muzic} with the possible exception of the Taurus Dark Cloud \citep{luhman09}.

Extending the work to massive clusters ($10^4$ M$_\odot$ or more)  confirmed the universality down to the limit of the observations of 0.5-1 M$_\odot$ \citep[e.g.][]{stolte06,andersen09,dario09}, where derived mass functions (MFs) in all cases were fit by  single power-laws. 
In order to claim the IMF non-universality, it would be crucial to observe that the flattening and peak position of the IMF are different in massive young clusters, nearby star forming regions, or the Galactic field.

It has been reported that the inferred IMF in the centre of massive elliptical galaxies is overabundant in low-mass stars (0.3 M$_\odot$ and below) compared to the  Galactic field IMF. 
The result is based on the strength of FeH and Na lines in the integrated spectrum of the galaxy centre. 
 \citet{conroy12,dokkum10} concluded that a bottom-heavy IMF was necessary to reproduce the spectra. 
It was further found that the higher the velocity dispersion of the early-type galaxy { a  more bottom-heavy IMF had to be adopted to reproduce the observed spectra \citep{cappellari}}.  

The  extragalactic studies are based on integrated properties of stellar populations with several assumptions concerning the dark matter content of the galaxies, star formation history, and the mass-to-light ratio.  
Only within the Galaxy and the Magellanic Clouds { can individual low-mass stars currently} be resolved and the IMF be determined directly through star counts  instead of depending on integrated properties. 
We are thus forced to observe the few Galactic analogues to the more distant starburst events which may have been prevalent in the earlier times in massive elliptical galaxies through their merging history.

Westerlund 1 (Wd1) was identified as a  cluster that is rich in giant stars  \citep{westerlund61,westerlund_87,piatti} and it  became clear with near-infrared observations that  it is   a supermassive nearby star cluster ({ several times} ~$10^4$ M$_\odot$). 
The distance to the cluster is estimated to be in  the range 3.7-5 kpc, based on near-infrared colour-magnitude diagrams \citep[4.0$\pm$0.2kpc]{gennaro}, HI observations \citep[3.9$\pm$0.6kpc]{kothes07}, the binary method \citep[3.7$\pm$0.6kpc]{koumpia}, and the high mass evolved stellar content \citep[5 kpc]{clark05}. 

The presence of evolved stars in the form of numerous red supergiants, Wolff-Rayet stars, yellow supergiants, and luminous blue variables \citep{clark05,negueruela} and the presence of a pulsar in the cluster \citep{muno} suggested that the cluster is at least 3 Myr old and \citet{clark05} argue for an age of 5 Myr based on the post-main sequence population. 
The presence and brightness  of pre-main sequence (PMS) stars indicate an age of 3-5 Myr, depending on the distance \citep{brandner,gennaro}. 
\citet{natkud} also suggest an age of 5 Myr based a proper motion selected sample of cluster members. 

 Previous work  based on the stellar content above 3.5 M$_\odot$ {  has}  suggested  a total cluster mass of some $50\times 10^3$ M$_\odot$ \citep{brandner,clark05}, by extrapolating the IMF down to the brown dwarf limit.  
This makes Wd1 the most massive young Galactic star cluster known. 
Wd1 thus provides an ideal opportunity to study a supermassive star cluster in detail and to probe its low-mass content such that { it can be used as a resolved template for  high-mass extragalactic clusters. } 

Here we present deep HST/WFC3 near-infrared   imaging of a 4.0\arcmin$\times$4\farcm1 region centred on Wd1 and observations of an adjacent control field. 
The main goal is to constrain the low-mass IMF in a starburst-like environment. 

The paper is structured as follows. 
In Section 2 we present the data and discuss their reduction and calibration. 
Section 3 presents the source detection, photometry, and artificial star experiments. 
The main results are presented in Section 4. The colour-magnitude diagrams are presented and the  field star contamination is discussed. 
We estimate the  foreground extinction, distance, age, and ellipticity of Wd1.  
In Section 5 the mass function as a function of cluster radius is derived. 
Evidence for mass segregation is discussed and the total mass of Wd1 is estimated through the direct detection of the low-mass content. 
Finally, we conclude in Section 6.

\section{Observations and data reduction} 

Wd1 was observed with the HST in cycle 17 (Proposal ID 11708, PI Andersen) using  the WFC3 IR channel. 
Observations were carried out in the \filter{F125W}, \filter{F139M}, and \filter{F160W} filters. 
The diffraction limited spatial resolution for the images are 0\farcs135, 0\farcs140, and 0\farcs151, respectively, and the pixel scale is 0\farcs13.  
{ A total area of 4\arcmin $\times$ 4\farcm1 was covered in a 2$\times$2 mosaic with small overlaps between each quadrant.} 
The centre of the mosaic is (RA,DEC)=(16:47:04.2,-45:50:42).
Each  tile in the mosaic was observed a total of 7 times in each filter with small offsets of up to  10\arcsec\, to correct for bad pixels, persistence effects and cosmic rays. 
 The total integration times for each tile were 2444 seconds in the \filter{F125W} filter, 6294 seconds in the \filter{F139M} filter, and 2094 seconds in the \filter{F160W} filter. 
Each individual frame was acquired in the STEP50 sampling up the ramp integration scheme integrating to step 13, 12, and 16, respectively,  for the three filters\footnote{The integration schemes are described in the data handbook  http://www.stsci.edu/hst/wfc3/documents/handbooks/currentDHB/toc.html}.

 The raw images were reduced using the STScI provided {\tt calwf3} pipeline in the {\tt pyraf} environment. 
The sampling-up-the-ramp nature of the data was utilised to get a large dynamic range by only using the readouts to the point where the detector saturates for the brighter stars.

We adopted to work on each tile in the mosaic individually instead of the full mosaic to avoid even small amounts of residual distortion to complicate the star detection and photometry in the overlap regions between each of the four quadrants. 
An additional advantage is that the integrity of the photometry can easily be compared in the overlap regions of the mosaic. 

\begin{figure*}
\centering
\includegraphics[width=11.3cm]{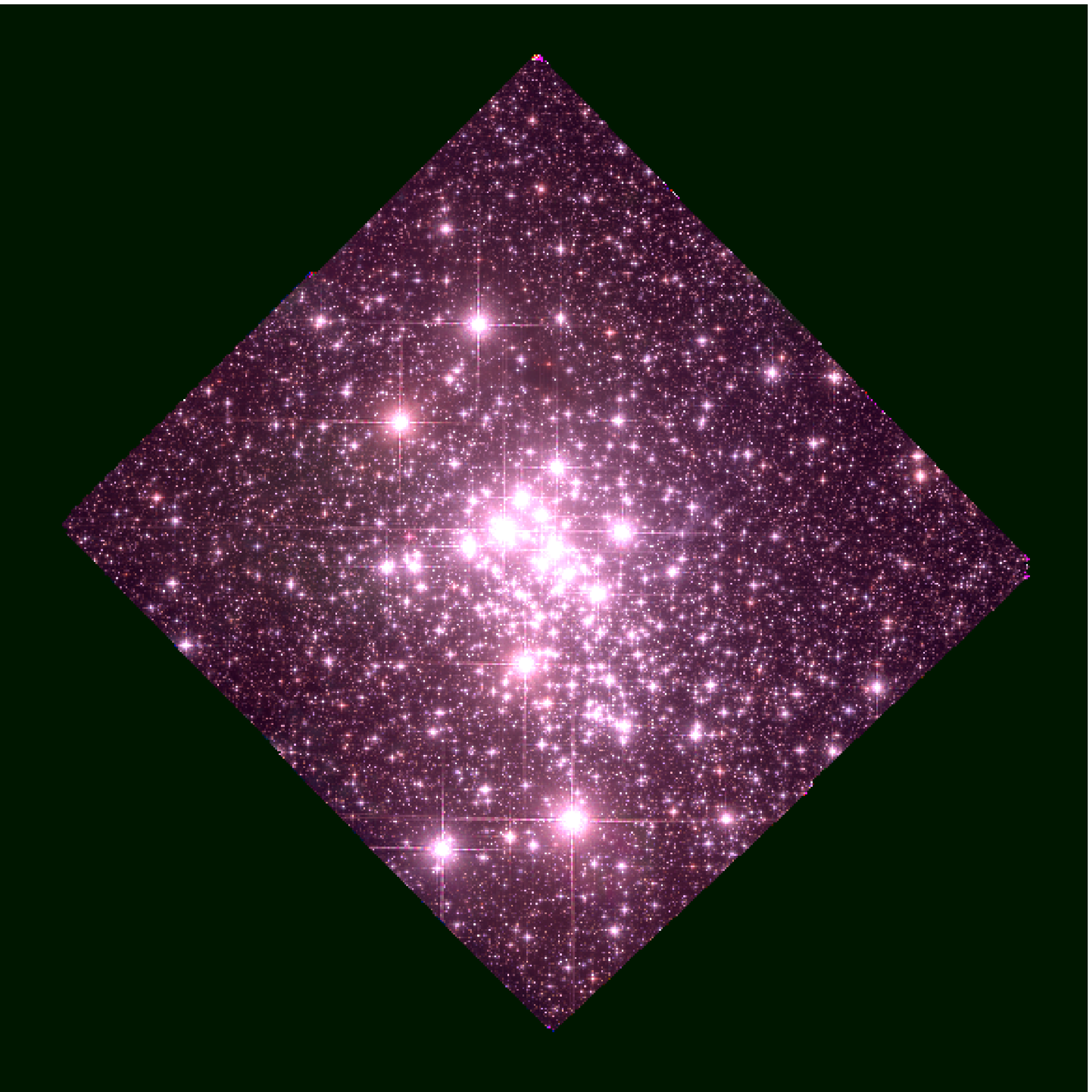}
\includegraphics[width=5.7cm]{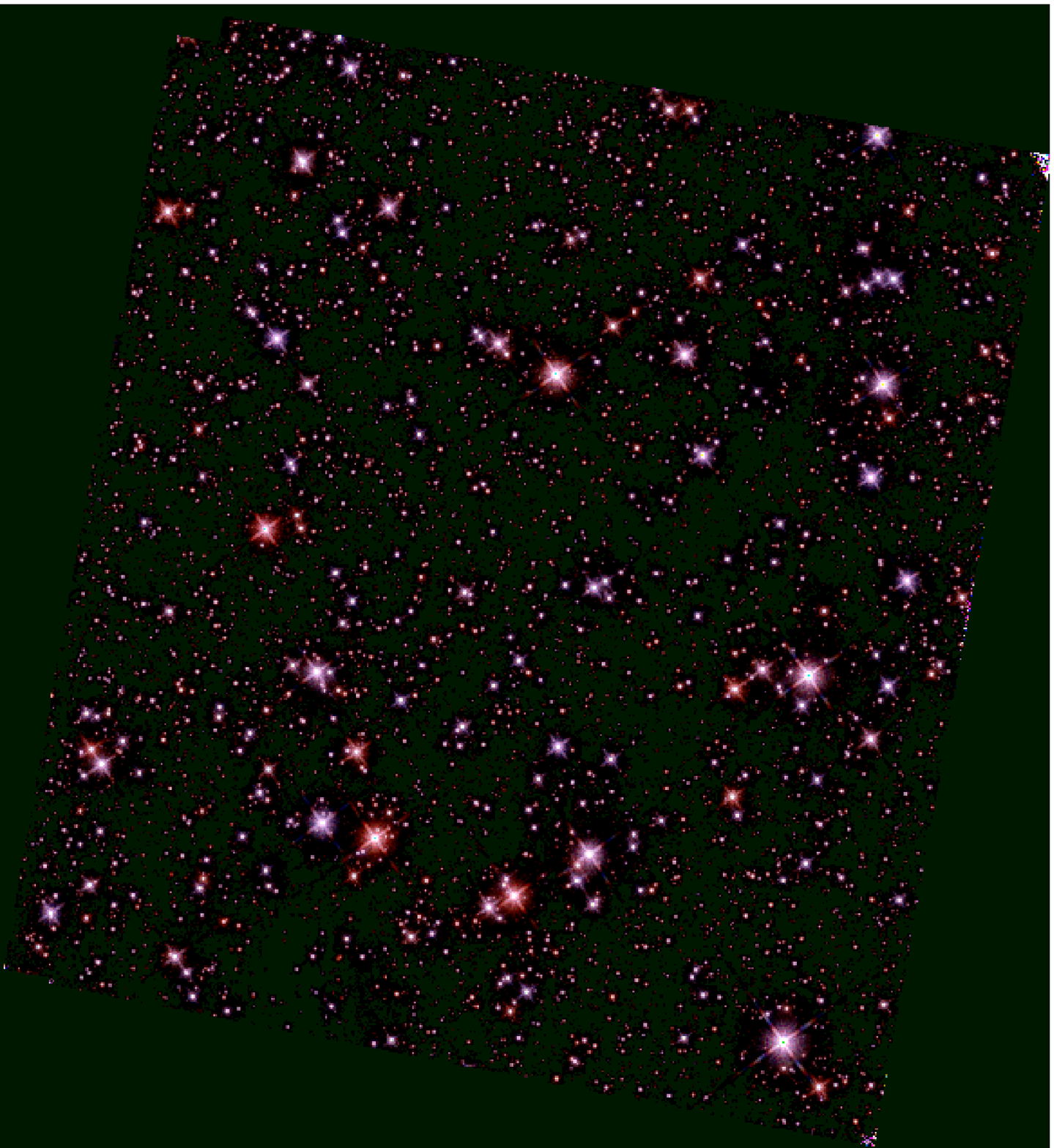}

\caption{Colour image of the HST observations of Westerlund 1 (left) and the control field (right). \filter{F125W} is the blue channel, \filter{F139M} the green, and \filter{F160W} the red channel. North is up and East to the left. The field of view of the cluster is 4\arcmin$\times4\farcm 1$ and 2\arcmin$\times2\farcm 2$ for the control field.  }
\label{mosaic}
\end{figure*}

 Offsets were refined using the iraf task  {\tt tweakshift} which is part of the STScI package and each  quadrant was combined utilizing the data quality flags provided as part of the STScI data products. 
Pixels saturated during the full exposure were  included as long as the first two reads were not saturated. 
The mosaic of the whole surveyed region is shown in Fig.~\ref{mosaic}. 

A control field was observed in addition to the Wd1 mosaic  in a similar manner  with the same total integration times for each filter.  
 A single pointing with small dithers was observed centred at (RA,DEC)=(16:47:42.787,-46:03:48.70) and the field of view is 136\arcsec$\times$123\arcsec.  
The position was chosen to be at the same Galactic latitude and close to Wd1 in Galactic longitude. 
The region has previously been used as a control field for more shallow observations of Wd1 \citep{brandner} and has been shown to possess similar line of sight reddening as Wd1. 
The data reduction was done in the same way as for the cluster tiles.

\section{Photometry and artificial star experiments} 
We present the source detection and photometry. 
The photometry is converted into the 2MASS system and compared with previous near-infrared photometry for Wd1.
The depth and completeness of the observations are characterised through artificial star experiments. 
\subsection{Source detection and photometry}
The source  surface density is very high  and we have opted for point-spread-function (PSF) photometry using the {\tt iraf daophot} package together with our own scripts. 
{ The main reasons  for the use of PSF photometry} are the ability to obtain photometry for sources even if they overlap and the ability to perform completeness analysis to quantify the effects of crowding. 
For each filter a first PSF model is created from the combined observed frames using bright non-saturated stars without any other bright stars within the PSF fitting radius. 
The stars were identified by hand and it was ensured that they were single objects through  visual inspection of their  profile. 
All objects, excluding the PSF stars, were then identified (see below) and removed from the frames using the PSF model. 
The PSF is then updated using the frames where all the other objects have been removed. 
These two steps are then repeated to improve the quality of the PSF model. 

The PSF radius for all filters is 17 pixels, corresponding to 2\farcs2, compared to a PSF full-width-at-half-maximum in the \filter{F160W} band of 0\farcs151.
The PSF  model is assumed  spatially constant and stars from all four quadrants were used for the PSF model. 
A total of 110 stars were used for the creation of the PSF. 

Source detection is complicated by the structure present in the PSF because of   spots in the diffraction rings caused mainly by the spiders holding the secondary mirror. 
Automatic source detection tends to identify the spots as stars which will result in spurious detections, especially at faint  magnitudes. 
To avoid this, we used a similar method as in \citet{andersen09} but expanded it  to take into account the availability of three filters. First the brightest non-saturated stars  are identified and removed with the PSF  model using {\tt allstar} to also remove the associated diffraction spots. 
Then the second brightest stars are detected, added to the list of the brightest stars and they are then removed from the original frame before fainter stars again are detected in the new star subtracted frame. 
This is repeated down to a detection limit of 4 times the background standard deviation. 
The \filter{F160W} observations are used for these steps of the source detection since they reach lower mass members  than the \filter{F125W} and \filter{F139M} observations because of the foreground extinction and { intrinsic red}  colours of low-mass PMS stars. 
The complete source list from the \filter{F160W}  observations is then used as the input file for photometry in the \filter{F125W} and \filter{F139M} filters. 
The centroid for each source is left as a free parameter when the PSF is fitted in the \filter{F125W} and \filter{F139M} bands. 

 With the position of each source observed through three filters it is now possible to identify spurious detections. 
 Any source found to move substantially (0.5 pixels) between the  filters is removed from the source list. 
The vetted source list is then re-run through {\tt allstar} again for the final photometry for all filters without the spurious detection. 
Objects that spatially align within 0.25 pixels are considered in the analysis. 
In total, 60561 sources  are detected in the cluster fields and 12117 in the control field.
The position, photometry, and completeness estimate (see below) for each source is provided in Table~\ref{sources_cluster}.
\begin{table*}
\caption{Astrometry, photometry and completeness for all sources in the field of Westerlund 1.}
\begin{tabular}{lllllllllll}
RA & DEC & X & Y & mag125 & merr125 & mag139 & merr139 & mag160 & merr160 & compl\\
251.73028 & -45.868391 & 2230.62 & 849.388 & 21.60 & 0.08 & 20.71 & 0.07 & 20.11 & 0.06 & 0.58\\
251.73729 & -45.868098 & 2093.65 & 857.652 & 17.95 & 0.12 & 17.47 & 0.13 & 16.99 & 0.15 & 0.93\\
251.73634 & -45.867928 & 2112.30 & 862.420 & 18.39 & 0.08 & 17.75 & 0.11 & 17.33 & 0.07 & 0.92\\
251.73243 & -45.867847 & 2188.76 & 864.673 & 21.29 & 0.10 & 20.72 & 0.12 & 20.29 & 0.12 & 0.61\\
251.73733 & -45.867726 & 2093.01 & 868.092 & 18.84 & 0.08 & 18.22 & 0.10 & 17.70 & 0.11 & 0.89\\
251.72990 & -45.867431 & 2238.10 & 876.317 & 20.88 & 0.11 & 20.11 & 0.07 & 19.55 & 0.07 & 0.76\\
251.72876 & -45.867324 & 2260.36 & 879.326 & 20.55 & 0.05 & 19.65 & 0.05 & 18.95 & 0.05 & 0.80\\
251.73000 & -45.867224 & 2236.14 & 882.122 & 20.91 & 0.06 & 20.14 & 0.09 & 19.51 & 0.09 & 0.74\\
251.73037 & -45.867218 & 2228.91 & 882.308 & 18.66 & 0.06 & 17.76 & 0.05 & 16.99 & 0.06 & 0.96\\
251.73064 & -45.867084 & 2223.75 & 886.060 & 19.12 & 0.07 & 18.25 & 0.05 & 17.54 & 0.04 & 0.94\\
\label{sources_cluster}
\end{tabular}
\end{table*}
\footnote{Table 1  is only fully available in electronic format the CDS via anonymous ftp to cdsarc.u-strasbg.fr (130.79.128.5)
or via http://cdsweb.u-strasbg.fr/cgi-bin/qcat?J/A+A/}

\begin{figure}
\centering
\includegraphics[width=8cm]{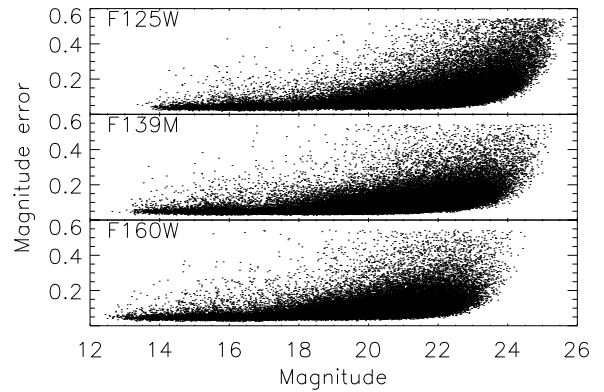}

\caption{The estimated magnitude errors as a function of derived magnitudes for all detected objects in the cluster fields.}
\label{mag_vs_merr}
\end{figure}

The depth of the photometry in the HST natural system is shown in Fig.~\ref{mag_vs_merr} where the magnitude errors are plotted as a function of derived magnitudes. 
Although objects are detected down to \filter{F160W}=24 mag, the completeness of the data is typically substantially lower and the cluster content can only be measured in a systematic way to brighter magnitudes. 
The completeness as a function of distance from the cluster centre is presented in Section~\ref{compl}. 
The minimum magnitude errors of 3\%\ are because of  the adopted uncertainty in the PSF moddeling including uncertainties based on  the different colours of the stars { implemented as a 3\%\ uncertainty in the flat field for the PSF fitting which is thus the minimum photometric error for each source.} 
The PSF model uncertainty was introduced in order to obtain average reduced chi square values of unity in the filters.

\begin{figure}
\centering
\includegraphics[width=8cm]{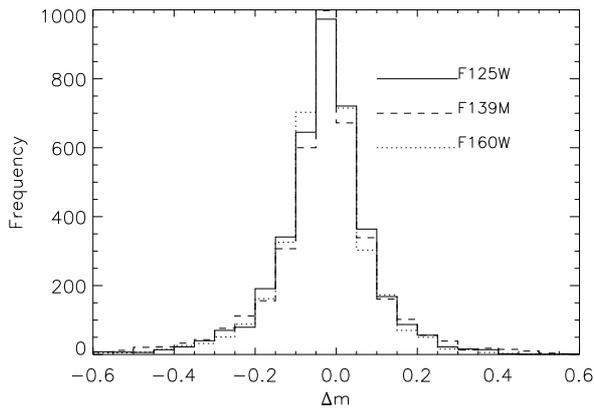}

\caption{A comparison of the photometry in the different filters for objects in the overlap regions of the mosaics. Only objects with formal photometric errors less than 0.1 mag are included.}
\label{int_integrity}
\end{figure}

The overlap regions in the mosaic can be used to test the internal integrity of the photometry. 
Fig.~\ref{int_integrity} shows the histogram of the magnitude difference for the stars located in two tiles for objects with formal errors less than 0.1 mag in each filter. 
A Gaussian fit to each histogram provides a fit with a peak within 0.02 mag of 0.00 mag and a standard deviation of 0.07 mag when constrained to the good quality photometry sources with formal errors less than 0.1 mag.
{ This is compared with the median predicted magnitude error from the PSF fitting of the stars in the overlap regions of 0.06 mag. }We conclude the photometry is robust despite the crowded nature of the observations, that there are no trends in the photometry between different frames, and that the photometric error estimate for the sources is reasonable.

\subsection{Conversion to the 2MASS system} 
 To convert the  photometry from the WFC3 system into the conventional 2MASS system we combined our data with the measurements of \citet{brandner} which were obtained with the SOfI instrument on the NTT  instead of 2MASS directly given the  large difference in depth and spatial resolution. 
The same region has been used as a control field for both this study and \citet{brandner} and the observations of the control field thus provides an excellent opportunity to obtain colour terms because of  the different amounts of reddening, and intrinsic colours,  for different stars. 

The SOfI dataset was further used to astrometrically calibrate the HST/IR data. 
Stars in common and brighter than \filter{H}=16.5 mag in the SOfI data were used for the astrometry calculation. 
Objects with their centroid  within 0\farcs2 between the SOfI and HST data were used for the astrometric calibration. 
A total of 485 
objects are selected, and the RMS of the angular displacement for this sample is 0\farcs44, half the seeing in the \citet{brandner} data.

\begin{figure}
\centering
\includegraphics[width=8cm]{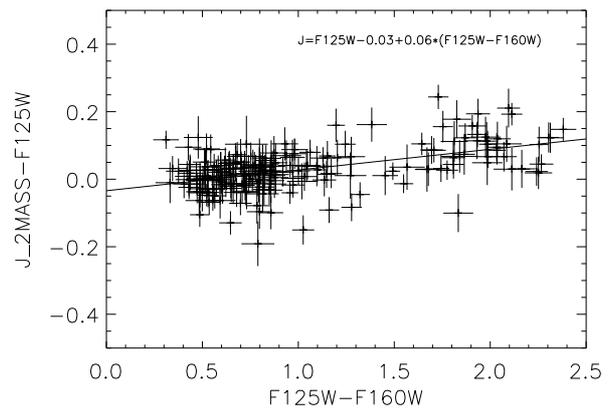}
\includegraphics[width=8cm]{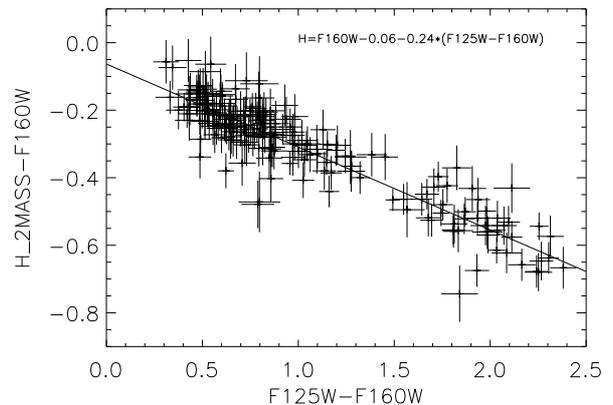}

\caption{Comparison of the \filter{J}-\filter{F125W} (left) and \filter{H}-\filter{F160W} (right) colours as a function of \filter{F125W}-\filter{F160W} colour for stars in common between the SOfI and WFC3/IR datasets. 
The best-fit first order polynomial for each dataset is overplotted.} 
\label{color_trans}
\end{figure}

 The SOfI photometry suffers  more from crowding effects than the current HST data due to the high surface density of stars even in the control field. 
We have therefore only used objects in common between the two datasets if { there are  no other  stars} within a radius of 20 WFC3/IR pixels  in the HST data within 1.5 mag in the \filter{F160W} band. 
Adopting objects with magnitude errors in the SOfI data of less than 0.05 mag in both the \filter{J} and \filter{H} bands restricts the sample to 187 
objects. 
Fig.~\ref{color_trans} shows  \filter{J}-\filter{F125W} and \filter{H}-\filter{F160W}  as a function of the \filter{F125W}-\filter{F160W} colour. 
{ Linear colour terms were determined by  Levenberg-Marquardt least-square fits and were found to be} 
\begin{eqnarray}
{J} & = & \filter{F125W}+a_{0J}+a_{1J}*(\filter{F125W}-\filter{F160W}) \\ 
{H} & =  & \filter{F160W}+a_{0H}+a_{1H}*(\filter{F125W}-\filter{F160W}),  
\end{eqnarray}
where the coefficients are given by $a_{0J}=-0.03\pm0.01$, $a_{1J}=0.06\pm0.01$, $a_{0H}=-0.06\pm0.01$, and $a_{1H}=-0.25\pm0.01$ { and  the HST photometry { is} in the VEGAMAG system.}  

The relatively large colour term for the \filter{F160W} band observations compared to other observations (e.g. NICMOS)  using the same filter is expected and is due to the short wavelength cutoff of the IR detector in WFC3. 

\begin{figure}
\centering
\includegraphics[width=8.5cm]{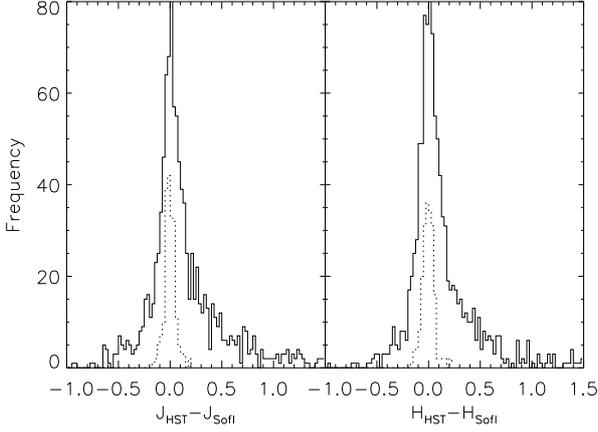}

\caption{The difference between the \filter{J} and \filter{H} SOfI and HST photometry for the control field, after both have been converted into the 2MASS system. 
The dashed-lined histograms show the differences for the objects identified as isolated objects  and the solid-lined histograms show the difference for all objects identified in both datasets. }
\label{us_vs_brandner}
\end{figure}

Fig.~\ref{us_vs_brandner} shows the histogram of the magnitude difference in both the \filter{J} and \filter{H} bands for all stars identified in both the SOfI dataset and the HST observations. The dashed-lined histograms are for the selected isolated objects used to determine the colour transformations whereas the solid-lined histograms are for all objects, i.e. including objects identified as potential blends in the SOfI observations based on the higher resolution HST data. 
{ The RMS for the high quality sample is 0.05 mag in the  \filter{J} band and 0.06 mag in the \filter{H} band.} 
The bright tail for the full sample is due to unresolved sources within the aperture in the lower resolution ground-based data that will systematically over-estimate the luminosity.

The cluster is plagued by saturated stars that are sufficiently bright that a region around them is heavily contaminated by the PSF wings from the bright star. 
We have identified the affected areas and masked them out in the analysis. All stars identified within the wings are ignored in the analysis and the area masked out is compensated for in the mass and number { density calculations}.

\subsection{Artificial star experiments}
\label{compl}
{  Artificial star experiments were carried out in order to quantify the effects of crowding and to determine the sensitivity of the data. 
{ We  followed an approach to derive multi-wavelength photometry that is an extension of that in \citet{andersen09}. }
Artificial stars with known magnitudes were placed in the frames with the PSF used for the photometry. 
We randomly chose 5\%\ of the stars observed, approximately 800 stars per quadrant per artificial star experiment. 
Artificial star were then placed in the frames with  the magnitudes and colours of the chosen stars.
The location of each star in the frames were determined  by adopting the same radial distance from the cluster centre as the observed star but at a random angle. 
This was done  to ensure the artificial star  follow a similar surface density profile as the cluster. }

\begin{figure}
\centering
\includegraphics[width=8cm]{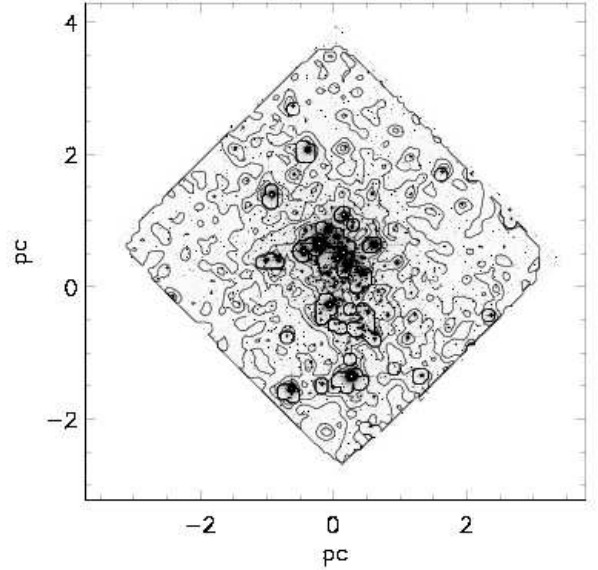}

\caption{Contour plot showing the 50\% completeness limit in the \filter{F160W} band as a contour plot to show both the general radial dependence of the completeness limit and the local effects due to bright stars. The spatial scale is assuming a distance of 4.0 kpc. Contours range from \filter{F160W}=15 mag to \filter{F160W}=22 mag in steps of 1 mag.} 
\label{recovered}
\end{figure}

{ The  detection and photometry processes were then performed with the same detection criteria as in the original frames. 
The process was repeated 400 times in order to obtain sufficient statistics for a two-dimensional completeness correction map such that radial symmetry does not have to be assumed. 
A total of 1.3 million artificial stars were placed in the frames.} 
An artificial star was considered recovered if its  detected  position was within 0.25 pixels of the input position, and the derived magnitude was within 0.5 mag of the input magnitude. 
The position accuracy is the same as the allowed difference for the objects stars between each filter. 
Fig.~\ref{recovered} shows a contour plot of the completeness in the \filter{F160W} band. 

\begin{figure}
\centering
\includegraphics[width=8cm]{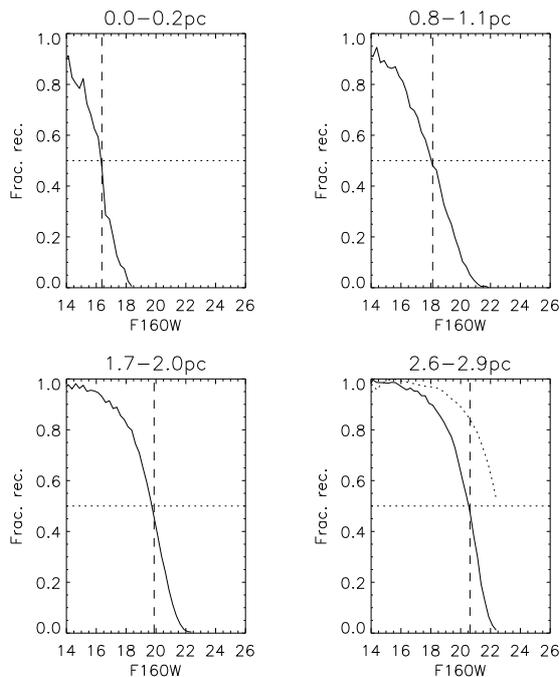}
\caption{{ The completeness fraction as a function of \filter{F160W} band magnitude for four different radial bins. The 50\%\ limit is indicated in each panel with the vertical dashed line. The completeness for the control field is shown in the lower right figure as the dotted line.} } 
\label{recoveredbins}
\end{figure}

{ The completeness is strongly dependent on the location within the cluster. } 
This is due to a combination of the local stellar surface density and the presence of bright stars. 
The very bright saturated stars make it difficult to detect any stars in their immediate vicinity. 
The crowding is  less { severe} at large radii where 50\% of the sources down to \filter{F160W}=20.9 mag are detected.

{ Figure~\ref{recoveredbins} shows the completeness as a function of \filter{F160W} band magnitudes for four radial bins in the cluster and the control field.
The control field is complete to fainter magnitudes as shown in the lower right panel. 
Although the crowding is similar to that in the outskirts of the cluster region, the lack of extremely bright stars  and the associated scattered light in the control field results in a 50\%\ completeness limit substantially deeper than for the cluster region. 
Because of  the rapidly variable completeness we use a local completeness correction for each object. 
Artificial stars within a 40 pixel radius are used to calculate the completeness correction for each stellar position. 
There are typically at least 1000 artificial stars detected within this radius.  

The input versus output magnitudes for the artificial stars have been compared. Fitting Gaussian profiles to the magnitude difference of the input to output magnitudes in each filter give a centre position to within 1\%\ (towards brighter retrieved magnitudes) and a $\sigma$ of the distributions of 0.06-0.07 mag. 
Slightly brighter retrieved magnitudes on average are to be expected due to crowding.
The widths of the distributions are almost identical to those derived for the photometry in the overlap regions of the mosaic. 
}

\section{Results}
The basic results from the colour-magnitude diagrams are presented. 
The subtraction of field stars is described and the basic cluster properties are { derived}.
Possible  ages and distances to the cluster are discussed.

\subsection{Colour-magnitude diagrams}
\begin{figure*}[!ht]
\centering
\includegraphics[width=13cm]{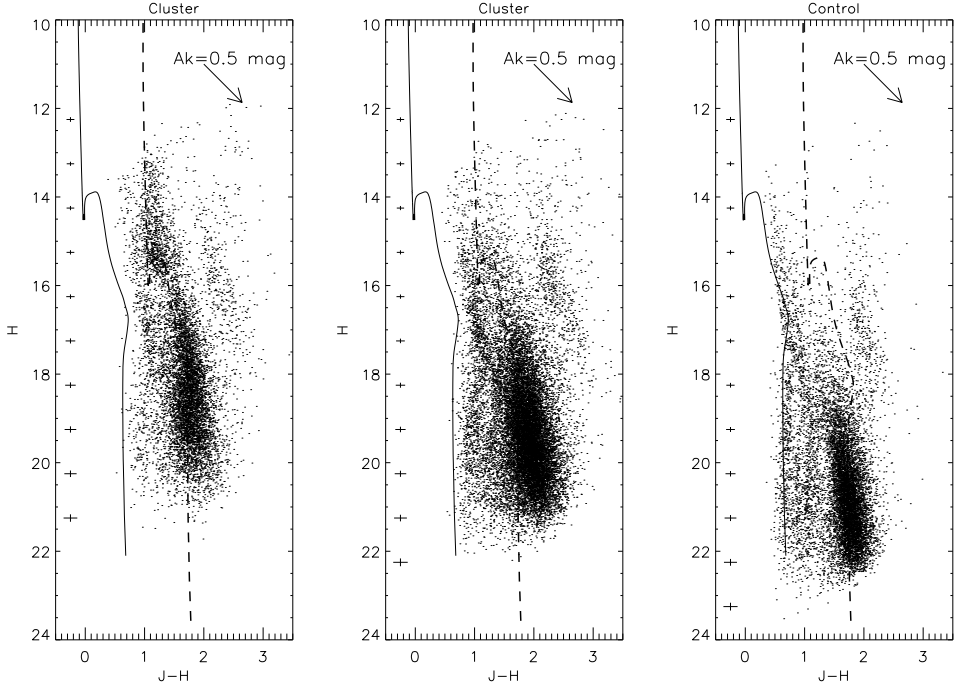}
\caption{{ Left: The  \filter{J}-\filter{H} versus \filter{H} CMDs for cluster members in an annulus  350-600 pixels (45\farcs5-78\arcsec, 0.9-1.5pc assuming a distance of 4.4 kpc) from the centre. 
Also shown are the reddening vector for the cluster field and an  un-reddened 4 Myr \citet{pallastahler} isochrone shifted to a distance of 4.4 kpc and reddened by A$_\mathrm{Ks}=0.87$ as determined in the text. 
The cluster main  sequence is  seen extending from \filter{J}-\filter{H}=0.8 mag  down to \filter{H}$=$ 16 mag, where the PMS population extends from \filter{J}-\filter{H}$=$1.3 mag  down to the depth of the data  of  \filter{H}=20.5-21 mag at this distance. 
Middle: the same but for stars more than 800 pixels (1.9pc) from the cluster centre. 
Right: The same colour-magnitude diagram for the control field covering an area 1.5 times that of the CMD to the left.} }
\label{CMD_1}
\end{figure*}

The main tool to characterise the cluster population and to separate cluster stars from the field star population is the \filter{J}-\filter{H} versus \filter{H}  colour-magnitude diagram (CMD). 
 { The location of the cluster population can be distinguished from the general Galactic population along the line of sight through a comparison of the cluster and control field CMDs. }
Fig.~\ref{CMD_1} shows the \filter{J}-\filter{H} versus \filter{H} CMDs for parts of the cluster field and the control field. 
The cluster main sequence (MS)  stands out in the diagrams compared to the field population, extending down to \filter{H}=16 mag. 
The same sequence was seen in the ground based near-infrared colour-magnitude diagrams \citep{brandner,gennaro,lim}. 
The hydrogen burning turn-on region, where the PMS merges with the MS,  is further seen at \filter{H}=16 mag. 
However, the  current dataset extends substantially deeper than the previous surveys and the PMS content can be seen as well. 

A comparison with the control field shows several regions of the colour-magnitude diagram where field stars contaminate the cluster population.  
A blue foreground population is seen extending from \filter{H}=14 mag, \filter{J}-\filter{H}=0.3 mag and extending down to the detection limit of \filter{H}=22 mag, progressively more red until \filter{J}-\filter{H}=1 mag. 
Field red clump stars are present at \filter{J}-\filter{H}$\approx$2 mag and \filter{H}=15-18 mag. 
Their location in the colour-magnitude diagrams for the cluster and control field are slightly different in the sense that  their colours are slightly bluer in the cluster CMD. 
The small difference in colour, and hence extinction, is likely due to small differences in the amount of extinction along the { two} line of sights and the low surface density of the objects. 
The last source of contamination is due to faint red objects extending in the control field 
from \filter{H}$\approx$18 mag to the detection limit with a colour \filter{J}-\filter{H}=1-2 mag.  
This population of likely background field stars  partly overlaps with the cluster PMS population in the CMD.

\subsection{Field object subtraction}
Although the location of some of the field stars is easy to take into account, there is an overlap between cluster objects and field objects in the colour-magnitude diagram for the fainter sources. 
The subtraction is performed using the \filter{F125W} and \filter{F160W} bands. 
Although the \filter{F139M} band could in principle be used in conjunction with the other two bands, we found that the magnitude difference was not sufficient to statistically provide more information since for almost all the sources their effective temperature is too  high to have strong water band absorption. 

{ We subtract the field stars in a statistical manner following the approach developed by \citet{gennaro}.
The densities of objects in the { CMD of}  the cluster  and the control fields { ($\rho_{on}$  and $\rho_{off}$, respectively)} are compared. 
The measurement errors are taken into account to provide a probability density function at the location in the CMD of the cluster field object and the corresponding density is calculated for the control field. 
 The completeness is lower for the cluster field than the control field which has to be taken into account as well. Each star in both the cluster and control fields is therefore weighted by its completeness when the surface density is calculated. 
Without this correction the faint objects closer to the cluster centre would be disproportionately subtracted. 
{ The ratio $(\rho_{on}-\rho_{off})/\rho_{on}$ is  then the probability a star is a cluster member.}

The method is illustrated in Fig.~\ref{field_sub}, where the probability of an object in the cluster field being  classified as a cluster member is colour-coded such that  red means a probability close to unity and violet a probability close to zero. 
\begin{figure}
\centering
\includegraphics[width=8cm]{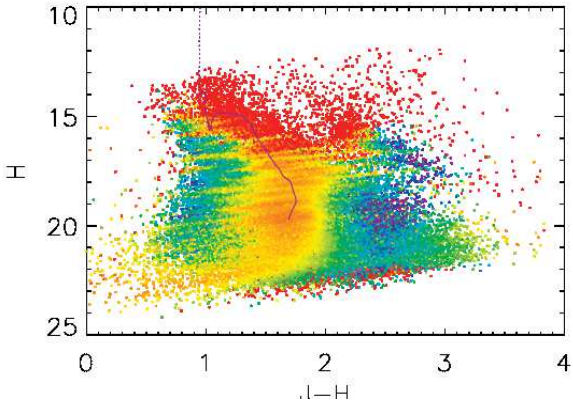}
\caption{\filter{J}-\filter{H} versus \filter{H} colour-magnitude diagram of Wd1. The colour scale { shows the probability for a star to be a} cluster member based on the field star subtraction method described in the text. 
  we note that the population of red clump stars erroneously characterised as high probability cluster members. 
This is due to the small differences in extinction along the line of sight.  
Overplotted is { a reddened (by A$_\mathrm{k}$=0.87)} \citet{siess} 4 Myr isochrone shifted to a distance of 4.8 kpc.  The { (reddened)} MS from \citet{marigo} is shows as the dotted line.}
\label{field_sub}
\end{figure}
{ The cluster  sequence is  recovered as high probability members, and the locus of high-probability members follows the 4 Myr \citet{pallastahler} isochrone  shifted by a distance of 4.4 kpc.
However, in regions of the CMD  with a low density of objects combined with high signal to noise the probability assignment works less well. }
This is most clearly seen for the region containing the red clump stars. 
They are formally high probability cluster members even though they are in fact expected to be background stars. 
This problem is caused by several factors. First, small differences in { the total} extinction along the cluster and control field lines of sight; second, small photometric errors; and third relatively small densities when compared to other regions of the CMDs. All these factors contribute to making the probability distribution non-smooth in the red clump region and the individual stellar membership probabilities less meaningful.
We will invoke an extinction limited sample (see below) to derive the cluster characteristics below and these objects will not be part of the cluster members once this selection has been applied where an object is considered a cluster member if its assigned probability is higher than a randomly number drawn from a uniform distribution. 

\subsection{Centre and ellipticity of Wd1}
Wd1 is not spherically symmetric as shown in \citet{gennaro}. 
{ We follow their approach to characterise the cluster profile using a generalised \citet{EFF} profile. } 
Briefly, the profile is defined by  a centre, the ellipticity of the cluster, a background surface density, the angle of the ellipse,  and a power-law surface density profile as a function of distance from the cluster centre, leading to  a total of 8 free parameters. 
{ Here the background surface density is set to zero since the field population is subtracted.} 

We derived the parameters only for  the regions of the image where the completeness is better than 50\%\  down  to \filter{H}=17 mag and a membership probability better than 75\%. 
Excluding regions with low completeness even for \filter{H}=17 mag in practice constrains the fit to a radius larger than 350 pixels (45\farcs5) and the fitting is extended to 700 pixels. 
{ The derived ellipticity is found to by 0.68-0.72 depending on the exact choice of inner and outer radius. 
\citet{gennaro} found the ellipticity to vary with  the mass range, extending from 0.55-0.8 with a tendency for higher ellipticity for a lower mass range, consistent with our findings. 
The angle of the ellipse is found to be 70-75 degrees clockwise where west is zero degrees. 
The centre of the ellipse is at (RA,DEC)=(16:47:04.1,-45:50:59).} 

\subsection{Derivation of stellar masses from isochrones}

{ No single set of stellar evolutionary models that include the PMS phase is available to  cover the total mass range of relevance for the dynamic range of the observations.} 
The  models by \citet{tognelli} cover the mass range 0.2-7 M$_\odot$ and thus a large fraction of the range covered by the observations. 
However, to reach the lower mass limit of the observations other tracks have to be used, for example \citet{siess}, \citet{pallastahler},  \citet{baraffe}. 
The latter covers fully the low-mass regime but lacks masses above 1.4 M$_\odot$. 
{ A  metallicity of Z=0.02 has been adopted for the \citet{siess} models, $[M/H]=0$ for the \citet{baraffe} and \citet{pallastahler}  models. The \citet{baraffe} models adopted have a mixing length of 1.0.}

The models from different groups can in principle be merged and one isochrone can be constructed for the whole mass range of interest. 
{  Since the isochrones from different groups differ in detail, merging of two isochrones will create artificial discontinuities at the merging mass which  will  persist in the derived MFs.} 
We have therefore derived the IMF for the different isochrones with the appropriate mass limits for each. 
Since the \citet{tognelli} tracks do not reach sufficiently low masses we have not used them in the analysis for the lower mass part of the IMF but we do include them in the discussion of the IMF when the content above 0.3 M$_\odot$ is considered. 
For the other isochrones, we utilise them over the { appropriate} mass ranges, i.e. \citet{baraffe} below 1.4 M$_\odot$ to the completeness limit of the data  and \citet{siess} in the range 0.1-7 M$_\odot$. 
For each selected age the \citet{pallastahler} isochrone covers the PMS phase. 
We thus use these tracks from 0.1-2 M$_\odot$.

{ The theoretical  effective temperature and bolometric luminosity for each object are converted into the 2MASS broad-band colours adopting  the BT-settle 2010  atmospheres \citep{allard} and utilizing the TA-DA software \citep{dario_tada} { since the atmosphere models are updated compared to the \citet{baraffe} published magnitudes.}  
The  atmosphere models predict too red colours when compared to observations for late spectral types \citep[e.g.][]{scandariato}. 
Colour corrections as a function of effective temperature were established for objects in the Orion Nebula Cluster, which has an age of 1.5-3.5 Myr \citep{reggiani,dario12}, for objects in the spectral class range K6-M8.5 \citep{scandariato}, covering the PMS stars in our sample.  
Although Wd1  is older than the ONC, the difference in surface gravity is very small for the objects on the PMS and  the corrections tabulated by \citet{scandariato} are suitable for  correcting the PMS stars in Wd1. 
A linear interpolation for the colour correction as a function of effective temperature was performed based on the effective temperature from the evolutionary models. }

\subsection{Foreground extinction}
\label{sec:ext}
Previous studies of Wd1 have determined a range of foreground extinction estimates. 
Near-infrared MS fitting  provides values of $\mathrm{A_K}=1.13\pm0.03$ \citep{brandner} and $\mathrm{A_K}=0.91\pm0.05$ \citep{gennaro}, depending on the adopted extinction law. 
Visual extinction estimates for the evolved stellar population have found values varying from $\mathrm{A_V=9.7\pm0.8}$ \citep{westerlund_87} and $\mathrm{A_V}=11.6-13.6$ \citep{clark05}. 
The optical and infrared estimates can be in agreement with each other assuming a ratio of total to selective extinction of 3.7 \citep{brandner}. 
An increase in $\mathrm{R_V}$ for higher optical depth through a molecular cloud has been observed for many sight lines \citep[e.g.][]{ascenso}. 
The extinction is not expected to be directly related to Wd1 but is due to  material along the line of sight.  
This is supported by the extinction  being very similar towards the control and cluster fields. 

{ Here we utilise the lower mass content to derive the foreground extinction. 
Only a very limited range of the MS is included in these  HST observations. 
Below ~1.5 M$_\odot$ the PMS isochrone is almost vertical in the near-infrared colour-magnitude diagrams and  de-reddening is possible with a unique solution for each star. 
The colour change for the PMS objects due to age variations within the expected age range of 3-5 Myr is  minor  since the effective temperature varies slowly and will not affect the derived extinction distribution. 
Adopting the \citet{baraffe} isochrones the \filter{J}-\filter{H} colour difference between the  3 Myr and 5 Myr isochrones is 0.01 mag or less, corresponding to a change in A$_\mathrm{Ks} < 0.01$. 
Since the isochrones are close to vertical there is little ambiguity due to distance uncertainties. 
 The extinction was derived for high probability members (probability of being a field star less than 20\%, mass below 1.4 M$_\odot$ and  photometric errors of less than 0.05 mag in each filter) and adopting a 4 Myr \cite{baraffe} isochrone. 
The extinction law of \citet{nishiyama} was adopted ($\mathrm{A_J}:{A_H}:{A_{ks}}$=2.89:1.62:1). }
We utilise the same approach as in \citet{andersen06} for objects without an infrared excess due to disks. 
The fraction  of objects with an excess detectable in the \filter{J} and \filter{H} bands is expected to be very low. 
The frequency is already modest for less dense clusters at the age of 3-5 Myr \citep{haisch,hernandez}. 
However, for the younger massive cluster NGC 3603 \citet{stolte06} showed { that the fraction is even lower} and already at 1 Myr only 20\% showed an \filter{L} band excess in the cluster centre. 
{ For typical circumstellar discs, we expect the fraction of stars exhibiting 
an excess in the H-band to be much lower than the fraction with an 
excess in the L-band \citep[e.g.][]{lada2000}.}
 
Further, the accretion rates derived  for the low-mass content in the ONC were found to be very low, interpreted as the disks being photo-evaporated at an early phase \citep{robberto04}. 
A population of stars with disks would skew the extinction distribution to higher values. 
The effect for weak lined T-Tauri stars is modest, \filter{J}=0.06 mag and \filter{H}=0.03 mag \citep{cieza}, corresponding to A$_\mathrm{Ks}$=0.02 mag.  

The distribution of derived extinction values is shown in Fig.~\ref{extinction}. 
\begin{figure}
\centering
\includegraphics[width=8.cm]{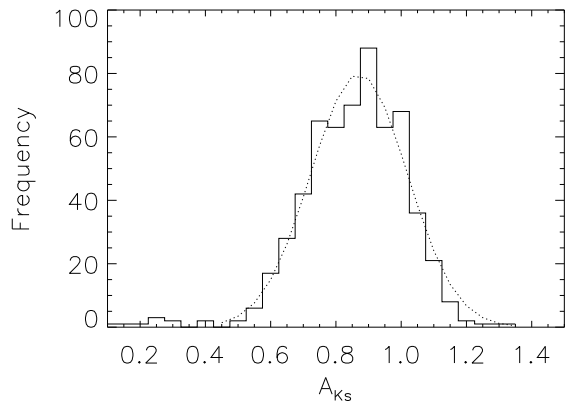}

\caption{Distribution of derived extinction for the PMS population, M$< 1.4$M$_\odot$ and with high photometric accuracy and membership probability. }
\label{extinction}
\end{figure}
The peak  of the extinction distribution is  A$_\mathrm{Ks}=0.87\pm0.01$ and  is  found from a Gaussian fit for extinction values less than A$_\mathrm{Ks}=1.3$ mag and greater than A$_\mathrm{Ks}=0.5$ mag.  The dispersion derived from the Gauss fit  is $0.14\pm0.01$ mag.   
The extinction law in { the  study of \citet{gennaro}} differs from the \citet{cardelli89} law in that it has a  steeper wavelength dependence on the extinction, thus explaining the lower extinction than in \citet{brandner}. 
We conclude the derived extinction is in agreement with the values determined from similar approaches with a similar extinction law. 

There is some evidence for variations of the extinction across the Wd1 region. 
This has previously been suggested \citep[e.g.][]{clark05} but only for relatively few evolved stars that tend to be close to the centre of the cluster. 
{ Fig.~\ref{ext_map} shows the extinction map for the cluster region based on the PMS population. 
This in turn limits our ability to determine the extinction in the central parts of the cluster.} 
\begin{figure}
\centering
\includegraphics[width=8cm]{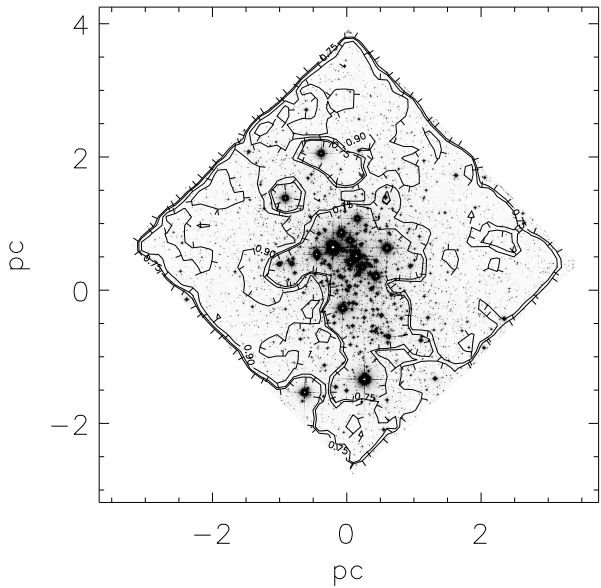}

\caption{An extinction map for Wd1 based on the PMS population. The  extinction is higher to the north and north-east. The hole in the centre is an artefact of the bright completeness limit that precluded the determination of  the extinction from the PMS stars. }
\label{ext_map}
\end{figure}
Although the extinction values only vary slowly across the cluster  there appears to be a slight excess compared to the average to the north and north-east of the cluster where  A$_\mathrm{Ks}$ approaches 0.95 mag. 
Thus, a constant extinction value cannot be used for the whole cluster. 
Instead, each object has to be de-reddened to the chosen isochrone along the reddening vector and an extinction limited sample applied before the MFs can determined. 

\subsection{Distance and age}
The age of Wd1 and its distance have been determined from both the post-main sequence content of the cluster and the MS to PMS transition.  
For the massive evolved stars \citet{clark05} argued from the different evolved stellar populations that the cluster age is between 3.5-5 Myr. 
This was  further constrained by \citet{crowther} to be 4.5-5 Myr using a larger sample of stars. 
 Utilizing the MS to PMS transition, \citet{brandner} suggested an age of 3.2 Myr and a distance of 3.55$\pm$0.17kpc. 
This was later revised by \citet{gennaro} to an age of 4$\pm$0.5 Myr and a distance of 4.0$\pm0.2$ kpc. 
\citet{natkud} suggested an age of 5 Myr, based on a proper motion selected sample of cluster members and an updated version of the \citet{tognelli} models. 
{ Cluster age and distance determined through the PMS hydrogen-burning  turn-on  are partly degenerate in the near-infrared CMDs,  in the sense that similar agreement between the data and the model can be obtained by a young distant cluster or an older cluster relatively nearby. }

An independent  distance estimate can be obtained  through HI absorption that traces the gas in the  spiral arms and then assumes Wd1 is located in the spiral arm \citep{kothes07}. 
{ The derived distance is 3.9$\pm$0.7 kpc,  based on the standard Galactic models of R$_\odot=7.6$ kpc and $V_\mathrm{rot}$=220 km/s. }

 The mass functions in Section 5 have been calculated assuming different ages and distances in the range suggested in the literature. 
Further, the age depends on the evolutionary models adopted. 
Thus, for a given cluster distance, different ages will be inferred when using different isochrones. 
We therefore derive the mass function for Wd1 adopting the combinations of distance and ages provided by the different isochrones.  
Since the age and distance are correlated we have employed isochrone ages of  3, 4, and 5 Myr and then  determined the corresponding distance that provides the best agreement between the model and the data  by visually aligning the model isochrone with the objects in the MS/PMS transition.
Table~\ref{table_dist_age} summarises the deduced distances for the three isochrones used, \citet{siess,pallastahler,baraffe}. 
For a given age this then provides the distance modulus. 
In the case of the \citet{baraffe} tracks where no MS-PMS transition is present due to the lower mass range covered the tracks were forced to match in the colour-magnitude diagram the 'kink' in the \citet{siess} isochrones around a mass of 1.4 M$_\odot$. 
The adopted distances range from 4.0 kpc in the case of the \citep{pallastahler} 5 Myr isochrone to 5.4 kpc in the case of the \citet{siess} 3 Myr isochrone (see Table~\ref{table_dist_age}). 

\begin{table}
\begin{tabular}{ccc}
Model & Age & distance\\
& (Myr) & (kpc) \\
\hline 
Siess et al. & 3 & 5.4\\
Siess et al. & 4 & 4.8\\
Siess et al. & 5 & 4.2\\
Baraffe et al. & 3 & 5.2\\
Baraffe et al. & 4 & 4.8 \\
Baraffe et al. & 5 & 4.4\\
Palla \& Stahler & 3 & 4.9\\
Palla \& Stahler & 4 &  4.4 \\
Palla \& Stahler  & 5 & 4.0 \\

\end{tabular}
\caption{Suitable distance and age combinations for the selected sets of  PMS tracks.}
\label{table_dist_age}
\end{table}

\section{Analysis}
We derive the  present day mass functions of Wd1 assuming different isochrones and distances an as a function of radius in the cluster. 
{ The  shape of} the low-mass part of the IMF is discussing through fitting the observed distributions with log-normal and power-law functional forms. 
The derived mass functions are further compared with previous results in the literature and the evidence for or against the universality of the underlying IMF is discussed. 
Finally, the total mass and the dynamical state of Wd1 are determined.

\subsection{Deriviation of the mass functions}
The completeness limit for the observations was shown in the previous sections to be a strong function of radius resulting in a limiting mass that  depends on the distance from the cluster centre. We therefore present the derived mass function at different radii { and only present fitting results down to the 50\%\ completeness mass limit for each radius}.
Due to the relatively large distance to Wd1  all binaries are unresolved. 
{ It is thus the system mass function assuming all stars are single stars that is derived in the following. }
The effect of binaries on the observed mass function have been examined in the past \citep[e.g.][]{kroupa91,dario09} and is found to depend  on the underlying single star mass function and the binary properties. 
Since the binary fraction in Wd1 is unknown, no good correction would be possible for the measured system mass function. 
Thus, comparisons with previously observed regions are performed between the  mass functions assuming all stars are single.

The higher mass limit is set by the saturation limit of the observations. 
The shortest read up the ramp for the WFC3/IR observations is  3 seconds, corresponding  to objects brighter than  \filter{J}$\sim12.5$ mag being saturated. 
However, this depends on the   background level which is  brighter in the  centre of Wd1 due to the scattered light from the very bright sources. 
This effectively lowers the dynamical range for the point sources.  
The brightness limit corresponds  to a (main sequence) star with a mass of  8 M$_\odot$. 
The \citet{siess}  models extend to 7 M$_\odot$ which is taken as the high mass limit. 

Each object is de-reddened to the isochrone { by } following the object back along the reddening vector and the mass is determined. 
The isochrones have been interpolated to substantially higher resolution in mass using the TA-DA software. 
The mass of each object  was  determined as the weighted average of the two nearest points on the interpolated isochrone.  
The differential extinction across Wd1 makes it necessary to de-redden each object individually to the isochrone instead of adopting a single value for the cluster. 

We identified the area of the CMD shown in Fig.~\ref{field_sub}  occupied by high-probability members, and used this area to define a confidence region for our final sample of members to be considered for the MF determination. 
We  adopted { an extinction limited sample from the} region of \filter{J-H} 0.35 mag on either side of the isochrone for the final sample. 
The limit corresponds to the width of high probability members (greater than 75\%) and the probability drops off rapidly thereafter. 
This sample contains roughly 80\% of the stars deemed members from the membership analysis. 
{ We have further performed the mass function fits with a more narrow distribution of \filter{J-H}=0.25 mag in order to test for any dependency of the results on the particular sample chosen. 
The fit parameters are shown in Fig.~\ref{fit_chab} as triangles.}

Fig.~\ref{mass_funcs} shows the mass functions at different radii derived for a \citet{siess} 4 Myr isochrone shifted to a  distance of 4.8 kpc and through one particular realisation of the field star subtraction. 
\begin{figure*}
\centering
\includegraphics[width=18cm]{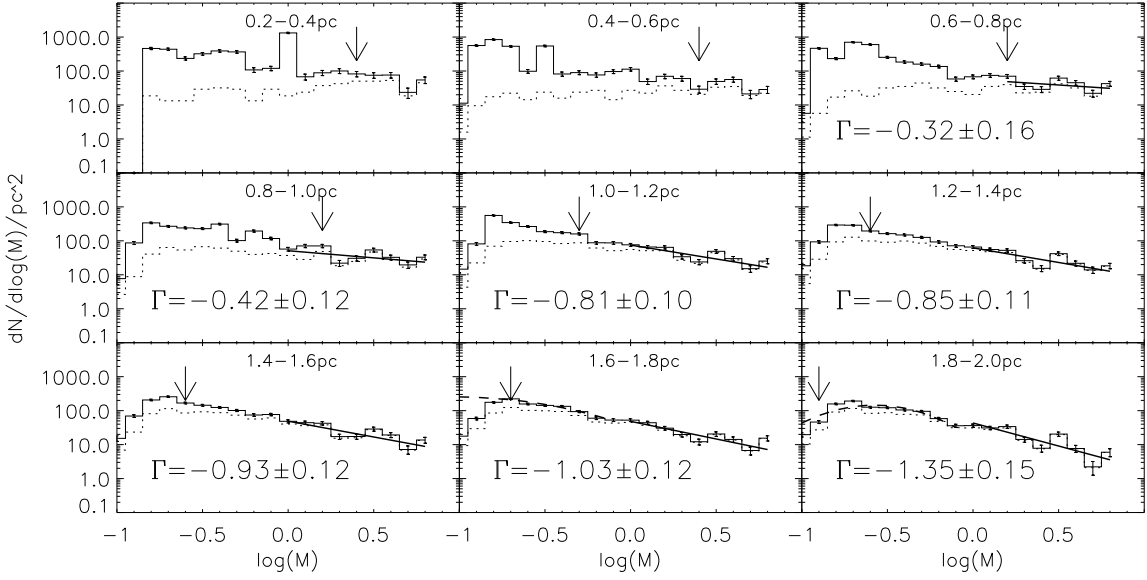}
\caption{Derived mass functions for Wd1 as a function of distance from the cluster centre for a \citet{siess} 4 Myr isochrone shifted to a distance of 4.8 kpc. 
For each annulus, the arrow indicates the 50\%\ completeness limit as determined in the artificial star experiments. 
The solid lined histograms illustrate the completeness corrected number counts whereas the dashed lined histograms illustrate the detected number of stars. 
Power-law fits are performed down to a mass limit of 1 M$_\odot$ and shown as solid lines.  } 
\label{mass_funcs}
\end{figure*}
The 50\% completeness limit is indicated with an arrow in each panel. 
Although the mass functions in general appear smooth there is a particularly strong feature at 2-4 M$_\odot$ for all radii. 
This is the transition region from the PMS to the MS for the age range 3-5 Myr. 
A comparison with the colour-magnitude diagrams shows that the isochrone is non-monotonic in this regime in magnitude. 
Similar features were  present in the mass functions presented by \citet{brandner} for Wd1, in NGC 3603 \citep{stolte06} and was also present in the luminosity functions for R136 in 30 Doradus \citep{andersen09}. 
Without an independent determination of the reddening or the spectral type of the objects in the transition region this degeneracy is difficult to overcome. 

{ Below  2 M$_\odot$ the mass functions derived from the \citet{siess} models are smooth and flattens at low masses outside  { 1.8 pc} where the 50\%\ completeness limit reaches 0.3 M$_\odot$ or less. }
For the models of \citet{baraffe} the situation is similar. 
However, although the isochrone extends to lower masses than the other, the low mass objects are less luminous for a given mass than for \citet{siess} resulting in a higher limiting mass. 
This is illustrated in Fig.~\ref{mass_funcs_baraffe} which shows the mass functions derived utilizing a 4 Myr \citet{baraffe} isochrone. 
\begin{figure*}
\centering
\includegraphics[width=18cm]{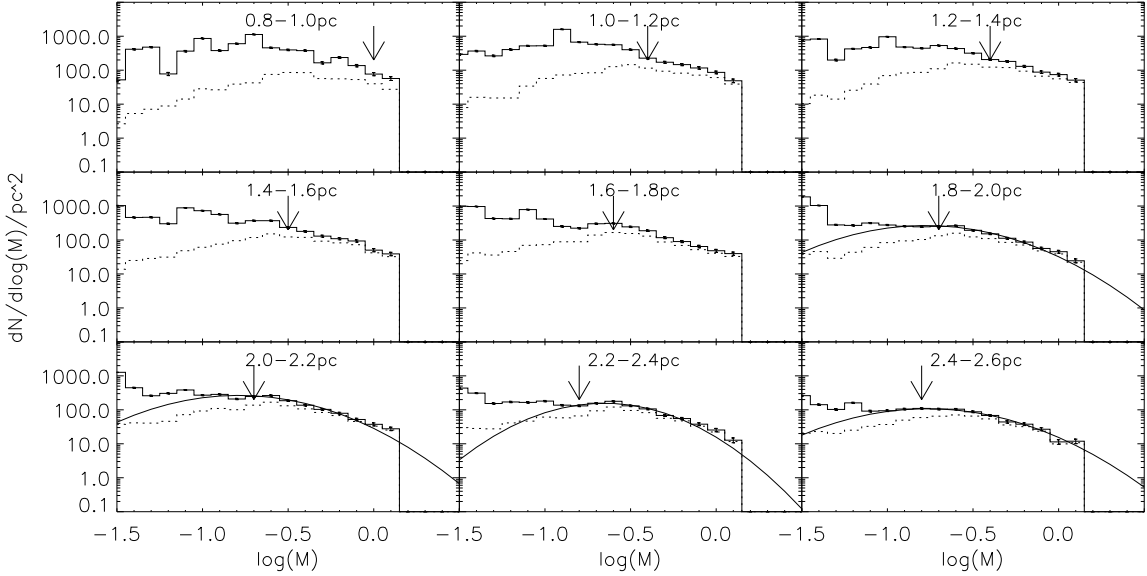}

\caption{Same as Fig.~\ref{mass_funcs} but here for the \citet{baraffe} 4 Myr isochrone and shifted to a distance of 4.8 kpc. 
The arrows indicate the 50\%\ completeness limit as determined in the artificial star experiments. 
Log-normal fits are performed below 1 M$_\odot$ down to the 50\% completeness limit and shown as solid lines.}
\label{mass_funcs_baraffe}
\end{figure*}

The following immediate conclusions can be drawn from the shape of the MFs. 
For the intermediate mass range, the MFs are for each radial bin well represented by a power-law but the slope appears to depend on radius. 
{ The  MF is found to flatten at large radii where the low-mass content can be determined.}
This is quantitatively very similar to what is observed for the Galactic field and in nearby star forming regions. 
Inspired by this we fit the derived mass functions with the functional forms commonly used in the literature, i.e. a log-normal distribution and segmented power laws. 

\subsection{Fits to the mass functions}
The field star subtraction is a stochastic approach in nature and the exact fitting results will depend on the specific realisation of the field star population (i.e. the seed of the random number generator). 
This is especially true  in regions with relatively low star counts per mass bin and where field star subtraction still occurs. 
{ Several field star subtractions were therefore performed in order to test the sensitivity to the specific realisation. 
200 individual field star subtraction realisations were performed with different seed values for the random number generator.}
Uncertainties to the parameters are determined from the distribution of fitted parameters. 
The  upper and lower $1/6$ quantiles are used as the error bars. 
Fig.~\ref{fit_chab} shows the results of the log-normal fits from the completeness limit up to 1 M$_\odot$ (the dividing point for the \citet{chabrier} IMF) for the different combinations of isochrone, ages, and distances. 
Power-law fits above 1 M$_\odot$ are shown in Fig.~\ref{fit_pow} where instead of the \citet{baraffe} isochrones the \citet{tognelli} isochrones are used. 
\begin{figure}
\centering
\includegraphics[width=8cm]{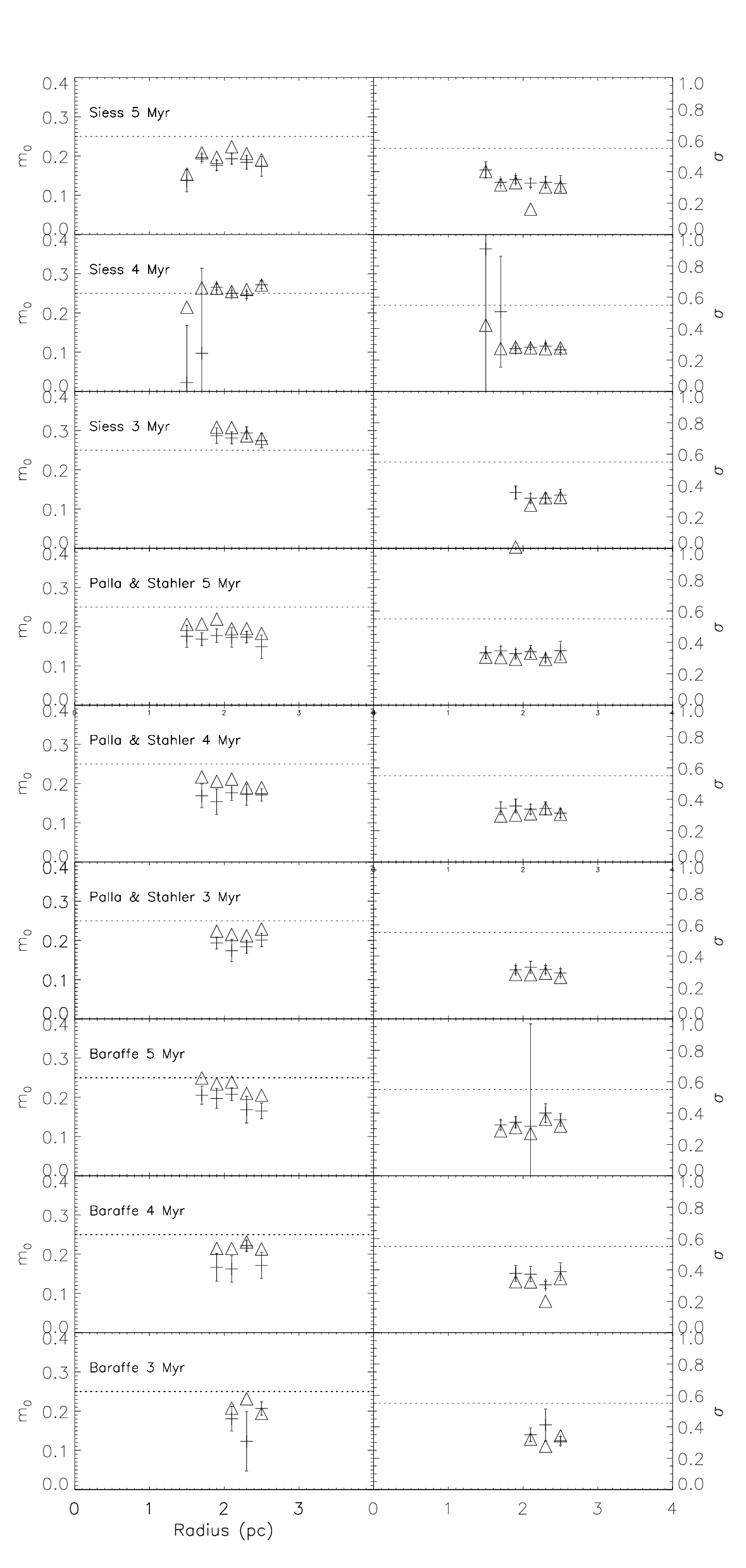}

\caption{Fit parameters $m_0$ and $\sigma$ for the log-normal fits to the low-mass part of the mass functions as a function of radius and for the three isochrones and  ages and up to 1 M$_\odot$. Indicated as dashed-dotted lines are the $m_0$ and $\sigma_0$ values prescribed in \citet{chabrier}.  
{ Plus signs are for the fits adopting the selection criteria of \filter{J-H}=0.35 mag whereas triangles are for \filter{J-H}=0.25 mag.}}
\label{fit_chab}
\end{figure}

Outside  a radius of  0.6-0.8 pc and depending on the assumed age of the cluster the 50\% completeness is at lower masses than the MS/PMS transition and a power-law can be fitted to histograms of the intermediate mass population. 
For  radii where the completeness limit { of 50\%} is at sufficiently low masses that the mass function histogram contains at least four bins below 1 M$_\odot$, the mass functions { were fitted with log-normal distributions $\frac{dn}{dlogm}\propto \exp(-\frac{(m-m_0)^2}{2\sigma^2})$, where for the Galactic field IMF \citet{chabrier} suggested $m_0=0.25$ M$_\odot$ and $\sigma=0.55$ M$_\odot$. 
Typical reduced chi square values for the fits are 1-3 depending both on the evolutionary models used  and on the adopted age.}
Within a radius of 1.5 pc the error bars are substantial for the subsolar population,  making it difficult to constrain the fit to the derived mass function. 

The derived { log-normal functional fits to the low-mass mass functions} outside 1.5 pc radius depends on the choice of isochrone and distance but is found to be on average 0.20-0.21 M$_\odot$ for the \citet{baraffe} isochrones, 0.20-0.27 M$_\odot$ for the \citet{siess} isochrones and 0.22-0.23 M$_\odot$ for the \citet{pallastahler} isochrones, where the range is for the three different age and distance combinations in each case. 
The spread in the width of the log-normal distribution is larger, and  
$\sigma$ is found to vary between 0.33-0.4 M$_\odot$ for the \citet{baraffe} isochrones, 0.29-0.55 M$_\odot$ for the \citet{siess} isochrones, and 0.28-0.48 M$_\odot$ for the \citet{pallastahler} isochrones. 

{ The log-normal fits will always provide a peak mass. We note that the difference between a two-segmented power-law as performed below and the log-normal fit are small in terms of the quality of the fits.
  This is further confirmed by the mass functions shown in Figs.~\ref{mass_funcs} and \ref{mass_funcs_baraffe}.
  Nevertheless, we can compare the derived log-normal fit parameters with those from other regions.
The log-normal fits provide in the outer parts of Wd1 a $m_0$ that is very close to the field IMF value, depending on the adopted isochrone. 
Comparing with the compilation in \citet{bastian} of young clusters, $m_0$ is within the range determined for other young star forming regions. 
They determined a range of $m_0$ between $-1 < \log m<-0.6$. }
The width of the derived mass functions are consistently more narrow than the value for the field of 0.55 M$_\odot$. 
A more narrow distribution was also found in for example the Pleiades \citep{moraux} when only fitted up to 1 M$_\odot$.

The log-normal fit of the mass function is just one possible functional form. 
Several others have been proposed, including segmented power-laws \citep[e.g.][]{kroupa_review}, a tapered power-law \citep{demarchi} or a heavy-tailed approximation to a log-normal distribution \citep{maschberger}. 
Fig.~\ref{fit_pow} shows the slope of a power-law fit above 1 M$_\odot$ for the \citet{tognelli} and \citet{siess} isochrones. 

\begin{figure}
\centering
\includegraphics[width=8cm]{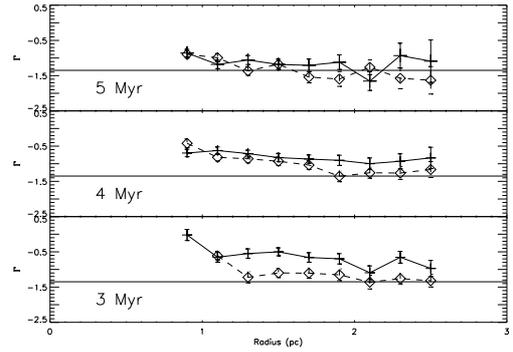}

\caption{Fit parameters for a power-law fit to the intermediate mass  part of the mass functions as a function of radius and for the \citet[][plus signs connected by solid line]{tognelli} and \citet[][diamonds connected by dashed line]{siess} isochrones for 3,4, and 5 Myr.  The fit is performed from 1-7 M$_\odot$. The horizontal dashed line is the Salpeter value  { of -1.35}. The different outer radii are due to the different distances to Wd1 for the different adopted ages. The value for a Salpeter slope is shown as the horizontal solid line.}
\label{fit_pow}
\end{figure}
The slope approaches a Salpeter value at larger radii from the cluster but there are indications of mass segregation in the inner parts of Wd1 in that the slopes are systematically more shallow than the  Salpeter slope. 
Similar results were obtained by { \citet{brandner}, using the \citet{pallastahler} pre-main sequence isochrones below 6 M$\odot$, and \citet{gennaro} for the higher mass content (3.4-27 M$_\odot$) where the \cite{dantona} pre-main sequence models were used}. 
\citet{brandner} found a shallow slope of $-0.6$ within 0.75 pc and a slope of $-1.3$ between 0.75 pc and 1.5 pc with a further steepening outside this radius. 
A similar steepening as a function of radius is found here although a Salpeter slope is only reached at 2 pc. 
This can partly be due to the slightly higher distances to Wd1 here than in the study of \citet{brandner}.

Power-law fits to the low-mass part of the IMF can also be performed. 
{ \citet{kroupa_review} suggests a change in slope of the IMF below 0.5 M$_\odot$ down to the brown dwarf limit. Similar limits were suggested for the fits to the  luminosity function for the central parts of the ONC \citep[][using the \citet{dantona} tracks]{muench}. }
Since the IMF derived by \citet{muench} is the unresolved mass function, we have used their break point mass of 0.6 M$_\odot$ ($\log m=-0.22$). 
The results are very similar for a break point mass of 0.5 M$_\odot$ ($\log m=-0.3$).
{ However, \citet{muench} obtained their fit to lower masses than possible here.
Based on the flatter slope for lower masses they observed we would therefore expect a steeper slope for a higher low-mass cutoff.
Fig.~\ref{pow_fits} shows the derived power-law slope for the \citet{siess} isochrones for fits above 0.6 M$_\odot$ and the \citet{baraffe} isochrones for the 0.15 $<$ M$_\odot < 0.6$ M$_\odot$ mass range. }

\begin{figure}
\centering
\includegraphics[width=7cm]{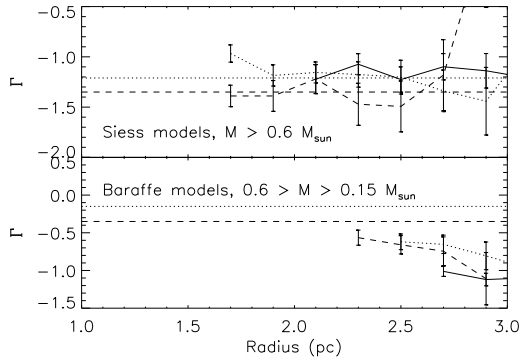}

\caption{{ Power-law fits for the low-mass content in Wd1. 
Fits are performed for masses larger than 0.6 M$_\odot$ (top, Siess isochrones) and for the mass range 0.15 M$_\odot$ $<$ 0.6 M$_\odot$ (bottom, Baraffe isochrones). The values derived for the ONC \citep{muench} are shown as the dotted lines and the \citet{kroupa_review} slopes are shown as the dashed line.}}
\label{pow_fits}
\end{figure}

{ The weighted average slopes for the \citet{siess} isochrones for M$_\odot > $ 0.6 M$_\odot$ up to 7 M$_\odot$ are $\Gamma=-1.0\pm0.1$. 
Similarly, using the isochrones from \citet{tognelli} provides a weighted average  of $-0.8\pm0.2$ for   M$_\odot > $ 0.6 M$_\odot$. 
The slope for the ONC was found to be  $-1.21\pm0.18$  \citep{muench}. 
For both surveys it is expected that very few, if any, binaries are sufficiently wide to be resolved and it is thus a direct comparison of the two  mass functions. }

Fits to the higher mass content in the cluster (above 3.4 M$_\odot$) have provided slightly steeper slopes. 
\citet{gennaro} found a global slope of $-1.44^{+0.08}_{-0.2}$ for the mass range 3.5-27 M$_\odot$. 
\citet{lim}, on the other hand, argue for a slope of $-0.8\pm0.1$ over the mass range of 5-100 M$_\odot$. 
Determining the mass of high-mass stars is difficult through optical and near-infrared photometry alone due to an uncertain bolometric correction. 
An unbiased spectroscopic sample would be necessary to address this in more detail.

{ For the lower mass content, 0.12 $< M <$ 0.6 M$_\odot$, the slope for the ONC was found to be $-0.15\pm0.17$ \citep{muench}.
  The data here does not reach the same lower mass limit but is restricted to masses above 0.15 M$_\odot$. 
The weighted averages for the $0.15$ < M$_\odot$ $< 0.6$ M$_\odot$  are $-0.5\pm0.2$, $-1.0\pm0.4$, and $-0.9\pm0.2$ for the \citet{siess}, \citet{pallastahler}, and \citet{baraffe} isochrones, respectively.
Depending on the models the slopes derived at slightly steeper than the case for the ONC, which is, at least partly, due to the higher cut-off mass. 
There is no evidence for mass segregation below 1 M$_\odot$ as opposed to what was observed for the intermediate mass range only. 
It is worth noting though that due to the fainter lower limit for this sample the 50\%\ completeness { above 1 M$_\odot$} the test can only be done outside $\sim$1.5 pc which would make any segregation difficult to detect. }

{ Although slopes were derived over the mass ranges provided in \citet{muench} a different choice of break mass than found in the ONC may be appropriate  for Wd1. 
Fig.~\ref{MF_zoom} shows this for the mass function derived in the outer parts of the cluster based on the \citet{baraffe} 4 Myr isochrone. }
\begin{figure}
\centering
\includegraphics[width=8cm]{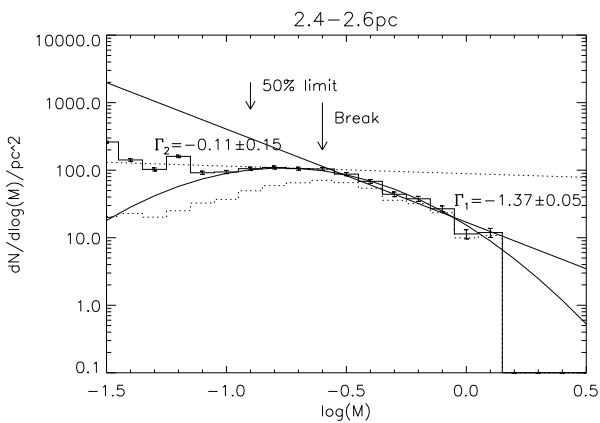}

\caption{A two segment power-law fit to the mass function at a distance of 2.4-2.6 pc based on a \citet{baraffe} 4 Myr isochrone. 
The mass where the slope has been chosen to change is $\log(M)=-0.5$. corresponding to 0.3 M$_\odot$. 
The best  log-normal fit is also shown.} 
\label{MF_zoom}
\end{figure}

{ A change of slope is more apparent at   $\log(M)=-0.5$ and the two power-law segments follow well the log-normal fit for exponents of $-1.37\pm0.05$ and $-0.1\pm0.2$ above and below the dividing mass, respectively. 

{ Common for many of the mass function fits in the literature is varying mass limits for the fits depending on the specific study.} 
This is particularly the case for the low-mass end where many of the log-normal fits in the literature for lower mass clusters and the field extend to much lower mass limits, often into the brown dwarf regime.} 
Furthermore, the fits are often performed on mass functions derived with different evolutionary tracks and different approaches to obtain the individual stellar  mass from broad-band photometry to spectroscopic classification. 

The studies of the Pleiades \citep{moraux} and the ONC \citep{dario12} use the same evolutionary tracks of \citet{baraffe} for the subsolar content. 
We have therefore performed  similar log-normal fits to their data over the same mass range as in this study. 
The parameters for the ONC are $m_0=0.20\pm0.06$ and $\sigma=0.58\pm0.13$ and for the Pleiades $m_0=0.28\pm0.04$ and $\sigma=0.45\pm0.09$. 
The values for the two clusters overlap with those determined for Wd1 using the same \citet{baraffe} evolutionary tracks and for the parameters fitted over the same mass range for the three clusters.

{ 
\subsection{Mass segregation in Wd1}
The systematic change in the slope of the mass function as a function of radius would suggest that the cluster is mass segregated for the mass ranges probed here. 
The higher mass content has been known to be mass segregated \citep[e.g.][]{brandner} but the present dataset allows to trace any mass segregation to lower masses. 
Fig.~\ref{cumu_mass} shows the cumulative mass functions for the outer radii adopting a \citet{tognelli} 4 Myr isochrone. 
\begin{figure}
\centering
\includegraphics[width=8cm]{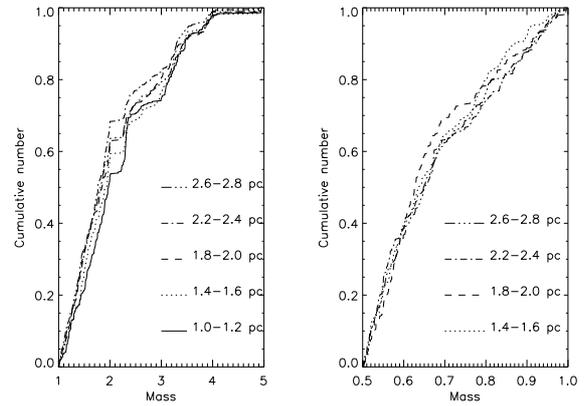}

\caption{Cumulative mass functions adopting a \citet{tognelli} 4 Myr isochrone.The  left panel shows { the cumulative mass functions} for the mass range 1-5 M$_\odot$, whereas the right panel shows { the cumulative mass functions} for the 0.5-2 M$_\odot$ range. The radial range where the comparison can be carried out varies for the two different lower mass limits. }
\label{cumu_mass}
\end{figure}
The left panel shows the cumulative distributions for the supersolar content outside a radius of 1 pc and after correction for incompleteness. 
There is a general deficit out lower mass stars closer to the centre, suggesting mass segregation. 
A two-sided KS test gives a probability that the mass functions outside 1.8 pc are drawn from the same underlying distribution as the mass function derived for 1.0-1.2 pc to be less than 0.6\%. 

The similar distributions for the content between 0.5 M$_\odot$ and 2.0 M$_\odot$ show a different picture. 
There is no systematic change { in the mass functions} as a function of radius as for the higher mass stars. 
Furthermore, the two sided KS test shows that the probability that the cumulative mass functions in the outer radii are drawn from the same parent distribution varies from 27\% in the 1.8-2.0 pc radial bin to 17\% in the 2.6-2.8 pc radial bin. 

Thus, the supersolar content of Wd1 appears mass segregated out to a radius of at least 1 pc. 
The subsolar mass content shows little evidence for mass segregation outside a radius of 1.4 pc where the completeness is sufficient to probe the subsolar mass function for mass segregation. }

\subsection{The mass of Wd1}
A calculation of the total present day mass of Wd1 is complicated by the crowding, especially in the central regions. 
Further, since these observations are saturated for objects more massive than 7 M$_\odot$, we are either restricted to extrapolation  or to use mass estimates from other studies. 

The total mass directly detected  from the field star subtracted sample  is found to be  15-20$\times 10^3$ M$_\odot$, depending  on the  realisation of the field star subtraction and on the adopted isochrone and distance. 
Taking the incompleteness corrections into account, for regions where the completeness is 50\% or better, this changes to 20-25$\times 10^3$ M$_\odot$. 

However, this is  a lower estimate of the total mass since only a fraction of the mass range is observed and the lower mass limit used depends on the location in the cluster. 
To estimate the full mass of the whole stellar IMF, up to an initial mass of  120 M$_\odot$, we have adopted the best fit functions for the IMF and then integrated across the whole mass range\footnote{Due to the steepness of the IMF for the high-mass end, the total mass limit only changes slightly with the upper mass limit. A higher upper mass limit for massive stars, as suggested in for example \citet{crowther_um} will therefore not change the conclusions.}. 
For the mass range above this survey we use the slopes of $-0.6$ for the central 0.75 pc, $-1.4$ out to 1.5 pc and $-1.6$ beyond in accordance with the evidence for mass segregation found in \citet{brandner}. 
Below 1 M$_\odot$, a log-normal distribution is assumed with $m_0=0.2$ M$_\odot$ and { $\sigma=0.33$} M$_\odot$, representative values from the mass function fits. 
For each annulus the total observed mass is determined for stars more massive than the 50\%\ completeness limit and this is corrected for incompleteness. 
Since only a fraction of the total mass range is observed this mass is then extrapolated to the whole stellar regime (0.08-120 M$_\odot$). 
Since there is no evidence for mass segregation at the low-mass part of the IMF we have adopted  the average fit parameters for the extrapolation across the whole cluster. 
The 50\%\ completeness limit in the inner 0.6 pc is above 1.4 M$_\odot$ and there is thus no mass estimate for the \citet{baraffe} and \citet{pallastahler} models that includes the inner parts of the cluster. 

{ The total mass out to a radius of 2.4 pc is found after extrapolating to the full mass range and including the stars within the extinction limited sample of \filter{J-H}=0.35 mag { and deemed as cluster members based on their membership probability}. 
Depending on the isochrone the mass is between $36-45\cdot 10^3$ M$_\odot$. 
However, this does not include the outer part of the cluster outside the surveyed area. 
\citet{brandner} estimated the mass outside 1.5 pc to be 33\%\ of the total mass. 
Calculating the total mass within 1.5 pc in our study and extrapolating with 33\%\ provide an estimate of $44-57\cdot 10^3$ M$_\odot$}

The determined masses agree well with those found previously for the higher mass content. 
\citet{brandner} estimated a total mass of 52$\times 10^3$ M$_\odot$ based on  objects in the mass range 3.4-30 M$_\odot$. 
\citet{gennaro} estimated a total mass of 49$\times 10^3$ M$_\odot$. 
The mass determined here is slightly lower than the 56$\times 10^3$ M$_\odot$ estimated from the massive stars  in \citet{clark05}. 
It should be noted that  this estimate is based on the evolved stars only and a directly observed mass of 1500 M$_\odot$. 
That the different mass estimates agree reasonably well is encouraging and gives confidence that the total mass of Wd1 is in the range 36-57$\times 10^3$ M$_\odot$. It is thus the most massive young Galactic star cluster known. 

The photometrically determined mass can be compared with the dynamical mass of the cluster. 
\citet{cottaar} determined the velocity dispersion of Wd1 through multi-epoch radial velocity measurements of a sample of the massive stars. 
They determined the velocity dispersion to be  $\sigma= 2.1^{+3.4}_{-0.9}$km/s (95\% confidence). 
Assuming a total mass of 50$\times 10^3$ M$_\odot$ and the cluster geometry determined by \citet{gennaro}, \citet{cottaar} also estimated that the velocity dispersion if Wd1 is in virial equilibrium should be 4.5 $\mathrm{kms^{-1}}$. They concluded that Wd1 is subvirial and bound. 
We confirm the findings on the total mass, and strengthen the conclusion the cluster is bound.

\subsection{Comparison with other young clusters and theoretical models}
{ For the first time is  the flattening  of the low-mass IMF in a supermassive cluster detected through direct star counts. 
{ The nominal  peak determined by assuming a log-normal fit is found outside 1.5 pc  to be $\sim$0.22 M$_\odot$, very close to the value found for the field IMF \citep{chabrier}. }
This also agrees with the the Pleiades \citep{lodieu12}, and the ONC \citep{slesnick} where the IMF was well sampled such that the detailed shape could be  determined. 
The widths of the log-normal fits are also comparable to the widths determined for both the Pleiades and the ONC, however they are all more narrow than what is found for the field ($\sigma \sim$0.55 M$_\odot$ for the field system IMF and $\sigma \sim$0.4 M$_\odot$ for the clusters). 
{We note that the  mass functions for Wd1 are still only being fit from the nominal peak mass of the log-normal fit  from the high-mass end. 
In contrast, the mass functions for the Pleiades and the ONC are determined beyond the brown dwarf boundary. } 
This combined with the current uncertainties in the pre-main sequence evolutionary tracks \citep[][e.g.]{stassun,hillenbrandwhite} this make the detailed comparison of the IMFs for different clusters difficult beyond comparisons of the general shapes as provided through functional fits.

Previously it has been shown that the ratio of brown dwarfs to subsolar stars for nearby embedded clusters, the Pleiades and the field IMFs are consistent with being drawn from the same underlying distribution \citep{andersen08}. 
Further, they were all consistent with being drawn from a log-normal distribution with the same peak mass and width as proposed by \citet{chabrier}, illustrating that the underlying distribution is { consistent with being} the same.

There have been several claims, from studies of extragalactic systems, that the IMF may vary as a function of environmental properties. 
Utilizing the Wing-Ford iron absorption feature, \citet{dokkum10} suggest that the underlying IMF in high velocity dispersion galaxies are systematically overabundant in low-mass objects compared to the Galactic field IMF. 
An IMF that not only continues to rise for late type objects but rises at least as fast as an extension of the { IMF to low masses with a Salpeter slope was suggested. }
A steepening of the IMF as a function of the velocity dispersion of elliptical galaxies has been found  through other independent studies, \citep{cappellari}. 
\citet{geha} on the other hand found for the ultra-faint galaxies Herc and Leo IV a flatter than Salpeter IMF for the mass interval  0.52-0.77 M$_\odot$ through star counts. 
Both { galaxies} have low metallicity and low velocity dispersion which appear to be in agreement with a general change of slope in the IMF as a function of  velocity dispersion of the host galaxy.
{ Such a flattening is not observed in the subvirial Wd1 cluster,     indicating that the MF is also determined from local star-forming     conditions}

We  conclude there is no strong evidence for a difference in the underlying IMF between Wd1 and other resolved star forming regions and the Galactic field. 
For all star forming regions where star counts have been possible the stellar IMF appears to be very similar. 
We note that small variations between regions can exist but may be hidden in the noise introduced from uncertain cluster parameters. 
 
Based on simple Jeans mass arguments one would expect variations in the IMF as a function of cluster density and temperature { of the star forming gas. }
However, more recent work including additional physics, in particular radiative feedback,  suggested the variations in the IMF should be modest \citep{bate09,krumholz}. 
\citet{hennebellechabrier} suggest the IMF originates from the clump mass spectrum and hence from compressible large-scale turbulence in clouds \citep[see also][]{padoan02}. 
This may thus indicate the very similar IMF derived for Galactic star forming regions is a result of the typical molecular cloud conditions in the Milky Way. 
Future observations of the brown dwarf content in Wd1 and of metal poor clusters in for example the Magellanic clouds will probe { the location of the peak of the IMF and whether it depends on metallicity and if the brown dwarf IMF is the same in massive clusters. }

\section{Conclusions}
We have obtained deep HST WFC3 near-infrared imaging of a 4\arcmin $\times$4\farcm 1 region centred on Wd1 together with a control field. 
The colour-magnitude diagram shows an extended cluster sequence progressing deep into the PMS regime down to brown dwarfs. 
We performed completeness tests that showed in the outer part of the cluster that the photometry is complete to a 50\% level down to 0.15 M$_\odot$ for the  first time in a young massive star cluster. 
Adopting the age and distance range from the literature, we derive the mass functions as a function of radius for the cluster for several different PMS isochrones. 
{ The flattening of the low-mass end of the mass functions is  identified in Wd1 and the mass of the flattening is found to be similar to that of nearby lower mass star forming regions. }
A log-normal fit to the mass functions shows the width of the distribution of the subsolar population to be comparable to or slightly less to than that of the Galactic field ($\sigma\sim0.33-0.44$ for the \citet{baraffe} models). 
Similarly, power-law fits to the low-mass mass function provide values comparable to other star forming regions. 
A two-segment power-law fitted to the outer parts of the cluster provides slopes of $\Gamma=-1.32\pm0.06$ and $\Gamma=-0.25\pm0.10$ for the mass ranges 0.6-1.4 M$_\odot$ and 0.15-0.6 M$_\odot$. 
At the low-mass end of rich clusters, no strong variations in the mass functions have been found to date. 
The intermediate mass content (1-7 M$_\odot$) is found to be mass segregated out to a radius of 2 pc. 
There is no evidence for mass segregation for the lower mass content outside a radius of 1.5 pc where crowding is less severe.

The total mass of the cluster is found to be 36-57$\times 10^3$ M$_\odot$, depending on the adopted isochrones and distances, in agreement with previous estimates. 
Together with previous estimates of the stellar velocity dispersion  this suggests the cluster will remain bound. 
}

\begin{acknowledgements}
  We thank the anonymous referee for many useful comments and suggestions that improved the paper.
  Based on observations made with the NASA/ESA Hubble Space Telescope, obtained [from the Data Archive] at the Space Telescope Science Institute, which is operated by the Association of Universities for Research in Astronomy, Inc., under NASA contract NAS 5-26555. These observations are associated with program \#11708.
\end{acknowledgements}

\end{document}